\documentclass[12pt]{article}

\newif\iffigs\figstrue

\usepackage{epsfig,latexsym}
\usepackage{amsmath}
\usepackage{color}
\usepackage{verbatim}
\usepackage{mathrsfs}
\usepackage{amssymb}

\DeclareMathAlphabet{\mathpzc}{OT1}{pzc}{m}{it}

 \csname
@addtoreset\endcsname{equation}{section}

\def\gz0{\gamma^{0}}

\def\a{\alpha}

\def\beq{\begin{equation}}
\newcommand{\eeq}[1]{\label{#1}\end{equation}}
\def\bea{\begin{eqnarray}}
\newcommand{\eea}[1]{\label{#1}\end{eqnarray}}
\def\ba{\begin{array}}
\def\ea{\end{array}}
\def\bec{\begin{center}}
\def\ec{\end{center}}
\def\ba{\begin{align}}
\def\ena{\end{align}}

\def\12{\frac{1}{2}}

\usepackage{slashed}

\usepackage{float}

\usepackage{booktabs}
\usepackage{caption}

\usepackage[mathscr]{euscript}

\usepackage{color}
\usepackage{graphicx}
\usepackage{epsf}
\usepackage{graphicx,epsfig}
\usepackage{amsmath}

\usepackage{multirow}

\usepackage{amsmath, amsthm, amssymb}

\usepackage{epsfig}
\usepackage{cite}
\usepackage{color,colordvi}

\newcounter{hran}

\makeatletter
\renewcommand\section{\@startsection {section}{1}{\z@}%
                               {-3.5ex \@plus -1ex \@minus -.2ex}%
                               {2.3ex \@plus.2ex}%
                               {\normalfont\large\bfseries}}
\makeatother

\vspace{0.5cm}

\newcommand{\bi}{\begin{itemize}}
\newcommand{\ei}{\end{itemize}}

\begin{document}

\begin{flushright}
CERN-PH-TH/2014-228\\
{\today}
\end{flushright}

\vspace{15pt}

\begin{center}


{\Large\sc $N=2$ Born--Infeld Attractors}\\


\vspace{35pt}
{\sc S.~Ferrara${}^{\; a,b,*}$, M.~Porrati${}^{\; c}$ and A.~Sagnotti${}^{\; a,**}$}\\[15pt]

{${}^a$\sl\small Department of Physics, CERN Theory Division\\
CH - 1211 Geneva 23, SWITZERLAND \\ }
\vspace{6pt}

{${}^b$\sl\small INFN - Laboratori Nazionali di Frascati \\
Via Enrico Fermi 40, I-00044 Frascati, ITALY}\vspace{6pt}

{${}^c$\sl\small CCPP, Department of Physics, NYU \\ 4 Washington Pl., New York NY 10003, USA}\vspace{6pt}

\vspace{8pt}

\vspace{35pt} {\sc\large Abstract}
\end{center}
\baselineskip=20pt
We derive new types of $U(1)^n$ Born--Infeld actions based on $N=2$ special geometry in four dimensions.  As in
the single vector multiplet $(n=1)$ case, the non--linear actions originate, in a
particular limit, from quadratic expressions in the Maxwell fields. The dynamics is encoded in a set of coefficients $d_{ABC}$ related to the third derivative of the holomorphic prepotential and in an $SU(2)$ triplet of $N=2$ Fayet--Iliopoulos charges, which must be suitably chosen to preserve a residual $N=1$ supersymmetry.
\vfill
\line(1,0){250}\\
{\footnotesize {$^*$On leave of absence from Department of Physics and Astronomy, U.C.L.A., Los Angeles CA
USA}}\\
{\footnotesize {$^{**}$On leave of absence from Scuola Normale Superiore and INFN, Piazza dei Cavalieri 7, 56126 Pisa ITALY}}

\noindent

\setcounter{page}{1}

\pagebreak

\newpage
\section{Introduction}
The supersymmetric Born--Infeld (BI) Lagrangian \cite{BI}, obtained in \cite{deser} and in a closed superspace form in~\cite{CF}, was shown in~\cite{BG} to encode the dominant low--energy couplings of the goldstino sector in the presence of a $2 \to 1$ partial breaking of supersymmetry. The original Volkov--Akulov~\cite{VA} action plays a similar role in the $1 \to 0$ case, so that this result can be also summarized by saying that in the $2 \to 1$ case the goldstino is accompanied by an $N=1$ partner, the Abelian vector field strength
\beq
W_\a \ =\ \bar{D}^{\,2} \, D_\a V \ .
\eeq{1}

The supersymmetric BI action possesses a number of special features. Clearly, setting to zero the gaugino it
reduces to the standard BI action for the vector field, while setting to zero the vector field it reduces to the
standard Volkov--Akulov action. Moreover, it is invariant under a second non--linearly realized supersymmetry, whose
 transformations can be conveniently expressed in terms of $W_\a$ and of the chiral superfield $X$, related to $W_\a$
 by the non--linear constraint~\cite{BG}
\beq
{W^2} \ + \ X\left( m\ - \ \frac{1}{4} \ \bar{D}^2 \bar{X}\right) \ = \ 0 \ .
\eeq{2}
Here $m$ is a parameter with dimension of $[\rm mass]^2$ and the additional supersymmetry transformations read
\bea
\delta W_\a &=& \left(m \ - \  \frac{1}{4} \ \bar{D}^2 \bar{X} \right) \eta_\a \ - \ i\ \partial_{\a \bar{\a}}X \ \bar{\eta}^{\bar{\a}} \ , \label{3} \\
\delta X &=& - \ 2 \ W^{\,\a} \, \eta_{\,\a} \ .
\eea{4}
Eq.~\eqref{2} can be regarded as a non--linear nilpotency constraint for an $N=2$ chiral superfield ${\cal X}$ \cite{RT}, which can be built by combining a pair of $N=1$ chiral superfields $X$ and $W_\a$ according to
\beq
{\cal X}(\theta_1,\theta_2) \ = \ X(\theta_1) \ - \ 2\, \theta^\a_2 \ W_\a(\theta_1) \ - \ \theta_2^{\,\a} \ \theta_{2\, \a} \left(m \ - \ \frac{1}{4} \ \bar{D}^2\bar{X}\right)\ .
\eeq{6}
The $N=2$ superfield ${\cal X}$ obeys the generalized superfield constraints~\cite{GSW,APT,FGP,RT}
\beq
D^2_{ab} \, {\cal X} \ = \ \epsilon_{ac} \, \epsilon_{bd} \, \bar{D}^{2\, cd} \, \bar{{\cal X}} \ + \ i\, m_{ab}\ , \qquad (a,b=1,2)
\eeq{7}
where $m_{ab}=\sigma_{ab}^x\,m_x$, and $m_{x}=(m,\, 0,\, 0)$ is a magnetic charge triplet. ${\cal X}$ also obeys
the nilpotency constraint \cite{RT}
\beq
{\cal X}^{\,2} \ = \ 0 \ .
\eeq{7a}
The solution of eq.~\eqref{7a} is provided by eq.~(\ref{2}), which also implies the additional constraints~\footnote{This corresponds to dropping $W_\a$ in eq.~\eqref{2} while keeping the nilpotency constraint in eq.~\eqref{8}.}
\beq
X^2\ =\ 0 \ , \qquad X\, W_\a \ =\ 0\ .
\eeq{8}
These are the $N=1$ nilpotency constraints proposed in~\cite{CDDFG,KS,R}.
Finally, the BI Lagrangian is simply
\beq
{\cal L} \ = \ \Im \ e \,  \int d^{\,2}\theta \ X \ .
\eeq{9}
Here $e$ is a complex parameter, $X$ is subject to the constraint (\ref{2}), while $\Re$ and $\Im$ will always denote real and imaginary parts.

Alternatively, in terms of ${\cal X}$, the Lagrangian becomes the half--integral of an $N=2$ Fayet--Iliopoulos term, projected on the $SU(2)$ triplet half--chiral measure (see~\cite{APT,RT,FGP}). The authors of~\cite{RT} also showed that the non--linear action~\eqref{9} can be obtained starting from the quadratic $N=2$ action considered by Antoniadis, Partouche and Taylor in~\cite{APT}. In that paper, the superpotential and the $N=1$ Fayet--Iliopoulos terms were chosen to give an $N=1$ vacuum with broken $N=2$ supersymmetry. A convenient way to obtain this result is via an electric charge $(e_1,e_2,0)$, aligned with the first two components of a triplet and a magnetic charge $(m,0,0)$, aligned with the first component. In this fashion, the first supersymmetry is unbroken and the $N=1$ Fayet--Iliopoulos terms vanish~\cite{RT}. On the other hand, the partial breaking $N=2 \to N=1$ is only possible if the $N=2$ Fayet--Iliopoulos magnetic charge $m$ does not vanish \cite{HLP}\cite{APT,FGP}.

Our goal here is to extend the construction to an arbitrary $N=2$ special geometry with $n$ vector multiplets, thus identifying the $U(1)^n$ generalization of eqs.~(\ref{2}) and (\ref{9}).

The model is defined by magnetic and electric charges, $m^A$ and $e_A$,
which will be defined in the next sections, and by the superpotential~\footnote{The cubic truncation leaves out
higher--order non--renormalizable terms that are expected to be subdominant at low energies. Interestingly,
this choice results in a shift symmetry of the axion fields $\Re \, X^A$.}
\beq
U(X)\ = \ \frac{i}{2} \ C_{AB} \, X^A \, X^B \ + \ \frac{1}{3!\,M} \ d_{ABC}\, X^A \,X^B \, X^C \ ,
\eeq{13}
where $C_{AB}$ and $d_{ABC}$ are totally symmetric and real and $M$ sets the scale of the problem.
For brevity, in the following we shall set $M=1$, keeping in mind that the dimensionless charge triplets
$Q_x=({m_x}^A,\,e_{x A})$  are meant to be accompanied by a factor $M^2$ in the final result.

We shall find it convenient to introduce shifted superfields $Y^A$,
$(A=1,\ldots,n)$, defined by
\beq
X^A \ = \ x^A \ + \ Y^A \ .
\eeq{10}
The vacuum expectation values (VEV)s $x^A = \langle X^A \rangle$ are determined by the $N=1$
vacuum condition
\beq
U_{AB}(x) \, m^B \ = \ e_A \ ,
\eeq{10a}
with
\beq
U_{AB}(X) \ = \ \frac{\partial^{\,2} U(X)}{\partial X^A\, \partial X^B}\ .
\eeq{12ab}

As we shall see, the $Y^A$ satisfy the generalized BI constraints
\beq
d_{ABC} \left[W^B\, W^C \ + \ Y^B\left(m^C \ - \ \bar{D}^2 \, \bar{Y}^C\right)\right]\ = \ 0 \ ,
\eeq{12}
which involve the totally symmetric sets of coefficients $d_{ABC}$ and reduce to eq.~(\ref{2}) for $n=1$, up to a slight change of conventions. As a result, the $U(1)^n$ generalized BI actions will depend on the choice of such symmetric tensors. We shall also examine in detail the available choices for the $d_{ABC}$ in the $n=2$ case. Moreover, we shall see that the $n$--extended Lagrangians can be cast in the form
\beq
{\cal L} \ = \ \Im \int d^2 \theta \ e_A \, Y^A \ - \ \Re \, \int d^2 \theta \ C_{AB} \left[ W^A\, W^B \ + \ Y^A \left( m^B \ - \ \bar{D}^2\, \bar{Y}^B \right) \right] \ ,
\eeq{12aa}
or alternatively, making use of the vacuum condition \eqref{10a} and of the non--linear constraint \eqref{12}, in the form
\beq
{\cal L} \ = \ - \ \Im \left[ U_{AB}(x) \int d^2 \theta \left( W^A\, W^B \ - \ Y^A \, \bar{D}^2\, \bar{Y}^B\right) \right] \ .
\eeq{12aaa}

Note that in the $n=1$ case the second term in eq.~\eqref{12aa} vanishes identically on account of the constraint \eqref{12}. This reflects the fact the single $C_{AB}$ that is present in that case can be eliminated by a field redefinition. However, for $n>1$ the $C_{AB}$ are needed, in general, to guarantee positivity, as is manifest from the alternative form of the Lagrangian in eq.~\eqref{12aaa}.

\section{Special Geometry, Fayet--Iliopoulos Terms and $N=1$ Attractors}

In this section we generalize the models of refs.~\cite{APT} and~\cite{RT} to the multi-field case. To this end, let us first observe that the data of the problem are the $N=2$ Fayet--Iliopoulos terms, which build up an $Sp(2n)$
symplectic triplet of electric and magnetic charges $Q_x\,=\,({m_x}^A,\,e_{x A})$, with $x=1,2,3$,  $A=1,..,n$, and the prepotential of eq.~\eqref{13}.

Eq.~(\ref{13}) clearly identifies the $d_{ABC}$ as third derivatives of the prepotential $U$. Moreover, the $N=2$ Lagrangian with an $N=2$ Fayet--Iliopoulos term, written in $N=1$ language, acquires a symplectic structure due to the underlying special geometry, which is encoded in the symplectic vector~\cite{STRO,CDFvP,SW,CDF}
\beq
{\cal V} \ = \ \left(X^A, \ U_A\, \equiv \, \frac{\partial U}{\partial X^A}\right) \ .
\eeq{15}
The scalar--field dependent $n\times n$ symmetric matrices $g_{AB}$ and $\theta_{AB}$ determine the quadratic terms in the vector fields as
\beq
{\cal L}\ = \ - \ \frac{1}{4} \ g_{AB}\, G_{\mu\nu}^A \, G^{B\, \mu\nu} \ +\  \frac{1}{8} \ \theta_{AB} \, G^A_{\mu\nu} \, G^B_{\rho\sigma} \ \epsilon^{\mu\nu\rho\sigma}\ .
\eeq{17}
Moreover, in $N=2$ special geometry
\beq
g_{AB}\ =\ \Im \,U_{AB}\ , \qquad \theta_{AB} \ =\ \Re \, U_{AB}\ , \qquad U_{AB}\ = \ \frac{\partial^{\,2} \, U}{\partial X^A \partial X^B} \ ,
\eeq{17a}
and it is convenient to define the symplectic metric
\beq
 \Omega \ = \ \left( \begin{matrix} 0 & -1 \\ 1 & 0 \end{matrix} \right) \ .
\eeq{17b}
The $2n\times 2n$ matrix ${\cal M}$, with entries
\beq
{\cal M} \ = \ \left( \begin{matrix} g \ + \ \theta \,g^{\,-1} \, \theta & -\,\theta \, g^{\,-1} \\ -\, g^{\,-1} \, \theta & g^{\,-1} \end{matrix} \right)\ ,
\eeq{17c}
then satisfies the two conditions of being symplectic and positive definite:
\beq
\qquad
{\cal M}\  =\  {\cal M}^{\,T} \ , \qquad {\cal M}\, \Omega \, {\cal M}\ = \ \Omega \ ,
\eeq{17d}
for a positive definite $g$, as required by the Lagrangian terms in eq.~\eqref{17}.

The contributions to the potential involve the triplets $Q_x=(m_x^A,\,e_{xA})$ of electric and magnetic charges. The first two combine into the complex sets
\beq
Q \ \equiv \ (m^A \,, \, e^A) \ =\ (m_1^A \ + \ i\, m_2^A\,, \, e_{\,1 A} \ + \ i\, e_{\,2 A})
\eeq{19}
and determine the superpotential
\beq
{\cal W} \ = \  {\cal V}^{\,T} \, \Omega \ Q \ = \ \left( U_A \ m^A \ - \ X^A \ e_A \right) \ .
\eeq{17e}
The last,
\beq
Q_3\ = \left(m_3^A\, ,\ e_{3 A}\right)\ ,
\eeq{19a}
is real and determines, in $N=1$ language, magnetic and electric Fayet--Iliopoulos $D$--terms.

The potential of the theory can thus be expressed, in $N=1$ language, as
\beq
V \ = \  V_F \ + \ V_D  \ , \
\eeq{19aa}
where
\bea
V_F &=& (\Im \, U^{-1} )^{AB} \ \frac{\partial \,{\cal W}}{\partial X^A}\ \frac{\partial \, {\cal W} }{\partial X^B} \  = \
\bar{Q}^T ({\cal M} \ - \ i\, \Omega ) Q \ , \label{21} \\
V_D &=& Q^T_3\, {\cal M} \, Q_3 \ .
\eea{22}

Vacua preserving an $N=1$ supersymmetry aligned with the $N=1$ superspace \cite{RT} are determined by critical points of the potential \footnote{Notice that this is not the case in the model of \cite{APT} where the $D$-term
has  a non--vanishing VEV, so that the unbroken supersymmetry is a mixture of the two original $N=2$ superspace supersymmetries \cite{PP}. The $N=2$ $SU(2)$ R--symmetry allows in fact to rotate $N=1$ Fayet--Iliopoulos terms into superpotential terms.}, and thus by the attractor equations
\beq
\frac{\partial \,{\cal W}}{\partial X^A}\ =\ 0 \ ,
\eeq{23}
which are in this case
\beq
({\cal M} \ - \ i\, \Omega)Q\ =\ 0\ .
\eeq{24}
This equation can admit a solution for nonzero $Q$ only if
\beq
i\, \bar{Q}^T\, \Omega \, Q \ = \ i\left( m^B\, \bar{e}_B \ - \ \bar{m}^B\, e_B \right) \ > \ 0 \ ,
\eeq{24a}
while the condition $V_D=0$ implies $Q_3=0$, since at the critical point ${\cal M}$ is positive definite. In solving the attractor equations we shall take $m^A$ real and $e_A$ complex, so that eq.~\eqref{24a} will translate into the condition
\beq
m^A\, {e}_{\,2\,A} \ > \ 0 \ .
\eeq{24b}

These are indeed attractor equations for the $N=2$ theory quadratic in vector field strengths. It is interesting to stress the analogy with the attractor equations for $N=2$ extremal black holes with symplectic
vector $Q=(m^A,e_A)$. In terms of the ${\cal M}$ matrix the black hole potential~\cite{FK,STRO2,FK2,CDF2,FGK},
\beq
V_{BH}\ =\ \frac{1}{2} \ Q^T {\cal M} \, Q \ ,
\eeq{24bb}
is also determined by the last expression in eq.~(\ref{21}), but for a real $Q$, so that the $\Omega$ term vanishes identically. However, in this case the value attained by $V_{BH}$ at the attractor point is positive and gives the Bekenstein--Hawking entropy
\beq
V_{BH}(X_{attr})\ = \ \frac{A}{4\, \pi} \ =\ \frac{S(Q)}{\pi} \ .
\eeq{25}
On the other hand, when expressed in terms of the central charge $Z$, which is the counterpart of ${\cal W}$, the black--hole potential contains an additional term~\cite{CDF2,FK,STRO2,FK2,FGK}, and reads
\beq
V_{BH} \ = \ |{\cal D}_i Z|^2 \ + \ |Z|^2\ .
\eeq{26}
Hence, at the $\frac{1}{2}$ -- BPS critical point, where ${\cal D}_i \, Z\,=\,0$,
\beq
V_{crit} \ \equiv \ V_{BH}(X_{attr}) \ = \ |Z|^2_{attr} \ .
\eeq{27}
Instead, in our case $V_{crit}=0$, which implies $\frac{\partial \, {\cal W}}{\partial X^A}=0$ in order to leave $N=1$ supersymmetry unbroken.

\section{Born--Infeld Attractors}

We can now exhibit a limit where the original theory quadratic in the field strengths gives rise to a generalized supersymmetric BI system, characterized by eqs.~(\ref{12}) and \eqref{12aa}. In $N=1$ language, the initial action reads
\beq
{\cal L} \ = \ - \ \Im \, \int d^2 \theta \, \left[U_{AB}\,W^A\,W^B \ + \ {\cal W}(X) \ + \ \frac{b}{2} \  \bar{D}^2 \, \left(X^A\, \bar{U}_A\ -\ \bar{X}^A\,  U_A\right)\right] \ ,
\eeq{29}
where
\beq
{\cal W}(X) \ = \ U_A \ m^A \ -\ X^A\ e_A \ .
\eeq{29a}
Therefore, the Euler--Lagrange equations for $X^A$ are
\beq
U_{ABC}\,W^B\,W^C \ +\ U_{AB}\, \left(m^B \ -\ b\,\bar{D}^2 \bar{X}^B\right) \ -\ e_A \ + \ b\, \bar{D}^2\,
\bar{U}_A\ =\ 0 \ .
\eeq{30}
The Lagrangian \eqref{29} is manifestly $N=1$ supersymmetry, while $N=2$ supersymmetry fixes the relative coefficients of the second and third terms. However, the coefficient of the superpotential can be changed by a rescaling of the complex charge vector $Q$, while the normalization $b$ of the scalar kinetic term reflects itself in the normalization of the $N=2$ supersymmetry anticommutator.

In the Introduction we have anticipated that $U(1)^n$ generalized BI Lagrangians can be defined via a set of $n$ constrained $N=2$
vector multiplets satisfying eq.~(\ref{12}). As we have stressed already, the parameters that enter the action combine into a constant, totally symmetric tensor $d_{ABC}$ and into a matrix $U_{AB}(x)$, which depends via eq.~\eqref{10a} on the $d_{ABC}$ and on the charge vector. The generalization of eqs.~\eqref{3} and \eqref{4} is
\bea
\delta \, W^A_\a &=& m^A\, \eta_\a \ - \ b\ \bar{D}^2 \, \bar{X}^A \, \eta_\a \ - \ i\, c\ \partial_{\a\bar{\a}} \, X^A \ {\bar{\eta}}^{\,\bar{\a}} \ , \label{3a} \\
\delta \, X^A &=& - \ 2\ W^{A \, \a} \ \eta_\a \ .
\eea{4a}
The closure of the supersymmetry algebra fixes the parameter $c=4\,b$, and in the following we shall choose $b=1$\footnote{If one demands, as in \cite{BG}, that the $N=2$ anticommutators have the same normalization, then $c=4\,b=1$.}.  Note that only the magnetic charges, and not the electric ones, enter the supersymmetry transformations. The reason is that the contribution to the superpotential ${\cal W}$ containing the electric charge is linear in $X^A$, and therefore is also invariant under the second supersymmetry \cite{BG}. Note also that the action~(\ref{29}) contains no other parameters.

The explicit form of the vacuum equations \eqref{23} and \eqref{24} is given in \eqref{5a} and implies that the goldstino is
\beq
\lambda_g \ = \
\left( \frac{i}{2}\ \bar{Q}^T \, \Omega\, Q \right)^{\,-\,\frac{1}{2}} m^{\,A} \ \Im \, U_{AB}(x) \ \lambda^B \ = \
\left( \frac{i}{2}\ \bar{Q}^T \, \Omega\, Q \right)^{\,-\,\frac{1}{2}} e_{2\,A} \, \lambda^{\,A} \ ;
\eeq{4aa}
therefore, the corresponding superfield takes the form
\beq
W_{g\,\a} \ = \ \left( \frac{i}{2}\ \bar{Q}^T \, \Omega\, Q \right)^{\,-\,\frac{1}{2}} e_{2\,A}\ W_\a^A \ .
\eeq{4aaa}
Its non--linear variation under the second supersymmetry, making use of eqs.~\eqref{24a} and \eqref{3a}, reads
\beq
\delta \, W_{g\,\a} \ = \ \left( \frac{i}{2}\ \bar{Q}^T \, \Omega\, Q \right)^{\,-\,\frac{1}{2}} e_{2\,A}\ m^A \ \eta_\a \ + \ \ldots \ = \ \left( \frac{i}{2}\ \bar{Q}^T \, \Omega\, Q \right)^{\,\frac{1}{2}} \ \eta_\a \ + \ \ldots \ ,
\eeq{4ab}
so that, in units of $M$, the supersymmetry breaking scale is
\beq
E \ = \ \left( \frac{i}{2}\ \bar{Q}^T \, \Omega\ Q \right)^{\,\frac{1}{4}} \ ,
\eeq{4abc}
which is a symplectic invariant, as expected.

Because of the nilpotency constraints on $X$, some care will be needed to obtain the non--linear actions of eqs.~\eqref{12aa} and \eqref{12aaa} from the spontaneously  broken theory of $n$ linear vector multiplets of Section 2. In particular, in order to satisfy the vacuum conditions (\ref{23}) it is necessary to introduce VEVs $\langle X^A\rangle =x^A \neq 0$. In fact, eq.~(\ref{24}) is
\beq
U_{AB}(x)\ m^B\  \equiv \ \left(i\,C_{AB}\ + \ d_{ABC}\ x^C\right)m^B\ = \ e_A\ , \quad m^B \mbox{ real ,  } \ e_A\ =\ e_{1 A} \ + \ i\,e_{2 A} \ ,
\eeq{5a}
and implies the two real equations
\beq
\left(C_{AB}\ + \ d_{ABC}\, \Im \, x^C\right)m^B\ =\ e_{2 A} \ , \qquad d_{ABC}\ \Re \,x^C\, m^B \ =\ e_{1 A} \ .
\eeq{6a}
A non--vanishing $C_{AB}$ is needed to restore positivity of the kinetic term when the matrix $d_{ABC}\, m^C$ is not positive definite.

If we now define chiral superfields $Y^A$ with vanishing VEV, letting $X^A=x^A + Y^A$, with the $x^A$ c--numbers, the equations of motion \eqref{30} become
\beq
d_{ABC}\left[ W^B\, W^C + Y^B\left( m^C - \bar{D}^2 \bar{Y}^C\right) \, + \, \frac{1}{2} \ \bar{D}^2 \left( \bar{Y}^B\, \bar{Y}^C\right)\right] \, + \, \left[\bar{U}_{AB}(x) - U_{AB}(x)\right]\bar{D}^2 \bar{Y}^B \, = \, 0 \ .
\eeq{30a}
Only the last term depends on $x^A$ (and also on $e_A$ via the vacuum equations \eqref{5a}).

The BI Lagrangians emerge in the limit in which $U_{AB}(x)$ is negligible with respect to the $d_{ABC}$, where the equations of motion reduce to
\beq
d_{ABC}\left[ W^B\, W^C + Y^B\left( m^C - \bar{D}^2 \bar{Y}^C\right) \, + \, \frac{1}{2} \ \bar{D}^2 \left( \bar{Y}^B\, \bar{Y}^C\right)\right] \, = \, 0 \ .
\eeq{m3}
The last contribution contains only overall derivatives, and therefore can be neglected in the IR limit where our effective actions will be well defined. One can then insert the \emph{ansatz }
\beq
U_{ABC}\, Y^B\, Y^C\, =\, 0
\eeq{deriv}
in~(\ref{m3}), solve the resulting equation
and check the self--consistency of the solution. This leads to the multi--field generalization of the BI constraint of \cite{BG,RT},
\beq
d_{ABC}\left[ W^B\, W^C + Y^B\left( m^C - \bar{D}^2 \bar{Y}^C\right)\right] \ = \ 0\ ,
\eeq{30c}
which was already presented in eq.~\eqref{12} in the Introduction. Taking into account that the equations of motion are solved by $D^A=0$, the $\theta^{\,2}$ component of \eqref{30c} reads
\beq
d_{ABC} \left[ G_{+}^B \cdot G_{+}^C \ + \  F^B \left( m^C \, -\, \bar{F}^C \right)  \right] \ = \ 0 \ ,
\eeq{30d}
where $G_{+}$ is the self--dual vector field strength and, here and in the next section, ``dots'' indicate full Lorentz contractions \footnote{The superfield expansion corresponding to our definition of $F^A$ is \ $X^A(\theta_1) = \ldots \ - \ \frac{1}{4} \ \theta_1^{\,\a} \ \theta_{1\, \a} \ F^A$.}

These complex algebraic equations determine the auxiliary fields $F^A$ as non--linear functions of $G_+\cdot G_+$ and $G_-\cdot G_-$, and are the seed of the generalized BI non--linear Lagrangians. For $n=1$ the Lagrangian corresponding to eq.~\eqref{12aa} reduce to the form
\beq
{\cal L} \, = \, \bigg( e_1 \, \Im \, F \ + \ e_2 \, \Re \, F \bigg) \, = \,
- \, \frac{e_1}{m} \ G \cdot \widetilde{G} \ + \ \frac{e_2\, m}{2}\, \left[ 1 \, - \, \sqrt{1 \, + \,
\frac{4}{m^2} \ G \cdot G \, - \, \frac{4}{m^4} \ \left(  G \cdot \widetilde{G} \right)^2}\, \right]\, .
\eeq{30e}

A simple way to verify that a field solving eq.~\eqref{30c} does indeed satisfy the ansatz \eqref{deriv} is to notice that the lowest component of~(\ref{30c}) is
\beq
d_{ABC}\left[ \lambda^B\, \lambda^C \ + \ y^{\,B}\left( m^C \, - \, \bar{F}^C\right)\right] \ = \ 0 \  ,
\eeq{m6}
where $\lambda^A=W^A|_{\theta=0}$ and $F^A$ is the auxiliary $\theta^{\,2}$ -- component of $Y^A$.
Multiplying eq.~(\ref{m6}) by $\lambda^A$ and using the Fierz identity $\lambda^{(A}\lambda^{B}\lambda^{C)}=0$ implies that

\beq
d_{ABC}\ \lambda^A\, y^B \left(m^C \ -\ 2\, \bar{F}^C\right)\ =\ 0 \ ,
\eeq{m61}
with $y^A=Y^A|_{\theta=0}$.
Since the factor within parentheses is arbitrary, this condition requires that
\beq
d_{ABC}\ \lambda^A\, y^B \ = \ 0 \ ,
\eeq{m62}
and multiplying eq.~(\ref{m6}) by $y^A$ and using \eqref{m62} one then finds
\beq
d_{ABC}\ y^A\, y^B \ = \ 0 \ .
\eeq{m622}

Therefore, eq.~\eqref{deriv} holds at $\theta=0$, and $N=1$ supersymmetry then implies that the entire multiplets vanishes. We have thus shown that the $Y^A$ obey the nilpotency equations
\beq
d_{ABC}\ Y^B\,Y^C\ = \ 0 \ , \qquad d_{ABC}\ Y^B \,W^C_\a\ =\ 0\ .
\eeq{30b}
Eq.~\eqref{30b} and eq.~\eqref{30c} combine in the $N=2$ superspace constraint
\beq
d_{ABC} \ {\cal X}^B \, {\cal X}^C \ = \ 0 \ ,
\eeq{m623}
which is the multi--field generalization of eq.~\eqref{7a} of the Introduction.

The Lagrangian corresponding to eqs.~\eqref{30a} is
\bea
{\cal L} &=& - \ \frac{1}{2\,i} \ \int d^{\,2} \theta \left[ U_{AB}(x)\, W^A\, W^B \, + \,
d_{ABC} \left( W^A \, W^B \, + \, \frac{1}{2} \ m^A \, Y^B \right) Y^C \right] \, + \, {\rm h.c.} \nonumber  \\
&-& \frac{1}{2\,i} \ \int d^{\,2} \theta \, d^{\,2} \bar{\theta} \left[ \bar{U}_{AB}(\bar{x}) \, Y^A\, \bar{Y}^B \, - \,  {U}_{AB}({x}) \, Y^A\, \bar{Y}^B \, + \, \frac{1}{2}\ d_{ABC} \left( Y^A\, \bar{Y}^B\,\bar{Y}^C \, - \, \bar{Y}^A \, Y^B \, Y^C \right) \right]\,  . \nonumber \\
\eea{m624}
Let us notice that the solution of eq.~(\ref{30a}) can be expressed as the solution of (\ref{m3}) with an additional term linear in $U_{AB}(x)$: $Y=Y|_{U_{AB}=0} \,+ \,\delta \,Y$. As a result $\delta \, Y \,= \, {\cal O}(1/\xi)$, where $\xi$ is an overall rescaling of the $d_{ABC}$, and after using the constraints \eqref{30c} and \eqref{30b} and some integrations by parts, one is then led in the $\xi\rightarrow\infty$ limit to the two equivalent Lagrangians presented in eqs.~\eqref{12aa} and \eqref{12aaa} of the Introduction.

Before concluding this section, we would like to comment on two aspects of
multi-field BI actions. First of all, let us emphasize some analogies and some
differences with the multi--field case considered in \cite{BMZ,ABMZ,RT,KT}. In those papers, the chiral superfield $X$ is a matrix, while in our case it is a vector. Moreover, their constraints are stronger. In fact, a $U(n)$ e.m. duality is imposed, while in our case we generally expect only a $U(1)^n$ duality even if the vectors are coupled.

Finally, we notice that our $U(1)^n$ construction is not a mere complexification of the construction in \cite{BG}, since for one matter it also applies for odd values of $n$. Moreover the terms containing $C_{AB}$ are crucial, in general, to grant positivity. This will be manifest in the simple examples that we are about to discuss, one of which could be related to the complexified nilpotency constraints
$(X\pm iY)^2=0$ in superspace. However, the corresponding action would contain ghosts unless a quadratic term involving $C_{AB}$ were added to the prepotential, and
this term necessarily breaks the complex structure.
Therefore, even in that particular case the model is different from the
$U(1)^{2n}$ generalizations proposed in~\cite{BMZ}.

\section{Explicit Examples: the $n=2$ Case}

The generalized BI Lagrangians are determined by the superfield constraints
\beq
d_{ABC} \left[W^A\,W^B \ +\ Y^B\left(m^C\ - \ \bar{D}^2\, \bar{Y}\right)\right]\ =\ 0\ .
\eeq{39}
To find them explicitly one needs only the F-term equations~(\ref{30d}), since the $D^A$-terms vanish.
Since eq.~(\ref{39}) is clearly solved by $F^A=0$ when $G^A_+=0$, it is useful to perform the change of variables
\beq
\Re F^A \ = \ \frac{1}{2} \ m^A \ - \ H^A \ ,
\eeq{40}
thus turning imaginary and real parts of eq.~(\ref{39}) into
\bea
d_{ABC} \ \Im \,F^B \, m^C &=& -\ d_{ABC} \ G^B\cdot \tilde{G}^C \label{41} \ , \\
d_{ABC}\, \left(\frac{1}{2} \ m^B \ + \ H^B  \right)\left(\frac{1}{2} \ m^C \ -\ H^C\right)&=& d_{ABC}\left(-\ G^B\cdot G^C \ + \ \Im F^B \,\Im F^C\right) \ .
\eea{42}
Notice that eqs.~\eqref{41} are linear, while eqs.~\eqref{42} are quadratic.

Any specific class of models solving these constraints is defined by the $U$ polynomial modulo field redefinitions by $Sl(n,R)$ transformations~\footnote{The more general symplectic duality $Sp(2n,R)$ is broken by the presence of the $N=2$ Fayet--Iliopoulos electric and magnetic charges
$(m_x^A,e_{x A})$.}. Inequivalent theories are thus classified by the $Sl(n,R)$ orbits of the cubic polynomials
\beq
U\ =\ \frac{1}{3!} \ d_{ABC}\, X^A\, X^B\, X^C\ .
\eeq{43}
As a first nontrivial example, let us consider the $n=2$ case, where the $d_{ABC}$, with $A,B,C=1,2$, take values in the
spin--$\frac{3}{2}$ representation of $Sl(2,R)$. This possesses a unique quartic invariant, which also corresponds to the discriminant of the cubic. The quartic invariant is
\beq
I_4\ =\ -\ 27 \,d_{222}^{\,2} \,d_{111}^{\,2} \ + \ d_{221}^{\,2} \, d_{112}^{\,2} \ + \ 18\, d_{222}\,d_{111}\, d_{112}\, d_{221} \ -\ 4 \, d_{111}\ d_{122}^{\,3}
\ -\ 4\, d_{222}\, d_{211}^{\,3}\ ,
\eeq{44}
and is a truncation of Cayley's hyperdeterminant, an object that also emerges from studies of black--hole entropies~\cite{KK,KL} and of q-bit entanglement in Quantum Information Theory~\cite{D,KL}. Different types of roots are associated to different properties of its four orbits: $O_{t},O_{s}, O_{l}, O_{c}$.

For $I_4>0$ the cubic has three real simple roots and $O_{t}$ is a \emph{time--like} orbit. When the roots are simple but two
are complex conjugates, $I_4<0$ and the orbit $O_{s}$ is \emph{space--like}. A double root $I_4=0$, $\partial I_4 \neq 0$
corresponds to a \emph{light--like} orbit $O_{l}$, and finally a triple root corresponds to $I_4=\partial I_4=0$ and to the critical
orbit $O_{c}$ made of a single point.

The four inequivalent theories can be associated to the four representative polynomials determined by the conditions
\beq
\begin{array}{lll}
I_4 \, >\, 0 \qquad & d_{222}\, =\, d_{211}\, \neq \, 0  \qquad & O_{t} \ , \\
I_4 \, < \, 0 &  d_{222}\, =\, d_{111}\, \neq \, 0 & O_{s} \ , \\
I_4\, =\, 0 & d_{222}\, =\, d_{221}\, \neq \, 0 & O_{l} \ ,\\
\partial \, I_4 \, =\, 0 & d_{222}\, \neq \, 0 & O_{c} \ ,
\end{array}
\eeq{45}
which read
\bea
&& O_t \ = \ \frac{1}{3!} \ X^{\,3} \ - \ \frac{1}{2} \ X \, Y^{\,2} \ , \\
&& O_s \ = \ \frac{1}{3!} \left(  X^{\,3} \ + Y^{\,3} \right) \ , \\
&& O_l \ = \ \frac{1}{3!} \ X^{\,3} \ - \ \frac{1}{2} \ X^{\,2} \, Y \ , \\
&& O_c \ = \ \frac{1}{3!} \ X^{\,3} \ .
\eea{45a}
The imaginary parts of the Hessian matrices of these polynomials contribute to the kinetic terms. It is simple to see that only in the $O_s$ case the Hessian is positive definite. On the other hand, the Hessians of the $O_t$ and $O_l$ cases have negative determinant, so that their eigenvalues have opposite signs. Finally, in the $O_c$ case there is a vanishing eigenvalue. Hence, aside from the $O_s$ case a $C_{AB}$ term is needed in the generalized BI Lagrangians.

We can now consider the solutions of the constraints given in  eqs.~\eqref{41} and \eqref{42}. The $O_{c}$ and $O_{s}$ cases are trivial, since there is no coupling between the two vectors in the non--linear constraints.
The other two cases are nontrivial and are determined by the nilpotency constraints
\bea
O_{t}: \ & X^2\ -\ Y^2  \ = \ 0\ , \qquad & X\, Y\ =\ 0\ , \nonumber \\
O_{l}: \ & X^2 \ =\ 0\ , \qquad & X\, Y\ =\ 0\ .
\eea{46}
Still, eqs.~(\ref{42}) can be solved by elementary techniques, since they only involve quadratic radicals.

For example, the explicit solution of eqs.~\eqref{41} for the $O_{t}$ case is
\beq
\Im F^X \ = \ \frac{m^X\,R^X \ + \ m^Y\,R^Y}{\left(m^X\right)^2 \ + \left(m^Y\right)^2}\ , \qquad \Im F^Y \ = \ \frac{-\ m^Y\,R^X \ + \ m^X\,R^Y}{\left(m^X\right)^2 \ + \left(m^Y\right)^2} \ ,
\eeq{50}
where
\beq
R^X \ = \ - \ G^X\cdot \widetilde{G}^X \ + \ G^Y\cdot \widetilde{G}^Y \ , \qquad R^Y \ = \ - \ 2\, G^X\cdot \widetilde{G}^Y \ .
\eeq{51}
On the other hand, eqs.~\eqref{42} become
\beq
- \left(H^X\right)^2 \ + \ \left(H^Y\right)^2 \ = \ S^X \ , \qquad 2\, H^X \, H^Y \ = \ S^Y \ ,
\eeq{52}
where
\beq
S^X \ = \ T^X \ - \ \frac{\left(m^X\right)^2}{4} \ + \ \frac{\left(m^Y\right)^2}{4} \ , \qquad  S^Y \ = \ T^Y \ + \ \frac{m^X\, m^Y}{2} \ ,
\eeq{53}
and
\beq
T^X \, = \, - \, G^X\cdot \,G^X \, + \, G^Y\cdot\, G^Y \, + \, (\Im F^X)^2 \, - \,  (\Im F^Y)^2 \ , \ T^Y \, = \, 2 \left( G^X\cdot G^Y \, + \, \Im F^X \, \Im F^Y \right) \ .
\eeq{54}
In terms of these quantities, the explicit solutions for $H^X$ and $H^Y$ read
\beq
H^X \, = \, \frac{1}{\sqrt{2}} \ \left( \sqrt{\left(S^X\right)^2 + \left(S^Y\right)^2} \ - \ S^X \right)^\frac{1}{2} \ , \ H^Y \, = \, \frac{1}{\sqrt{2}} \ \left( \sqrt{\left(S^X\right)^2 + \left(S^Y\right)^2} \ + \ S^X \right)^\frac{1}{2} \, .
\eeq{55}

The solutions of eqs.~(\ref{42}) become apparently more and more complicated with
increasing $n$, when the number of inequivalent cases and their degeneracies also increase. Their classification rests on the theory of invariant polynomials, which was only completed for the $n=3$ and $n=4$ cases so far \cite{inv_theory}.

\subsection*{Acknowledgements} We  are grateful to P.~Aschieri, A.~Corti, G.~Dvali, R.~Stora, V.~Varadarajan and A.~Yeranyan for useful discussions.
 S.~F. and A.~S. are supported by the ERC Advanced Investigator Grant n.~226455 (SUPERFIELDS). M.~P. is supported in part by NSF grant PHY-1316452. M.~P. and A.~S. would like to thank CERN for the kind hospitality and the ERC Advanced Investigator Grant n. 226455 for support while at CERN. A.~S. is also supported in part by Scuola Normale Superiore and by INFN (I.S. Stefi).

\end{document}